\documentclass[twocolumn,aps,prb,showpacs]{revtex4}

\usepackage{amsmath} 
\usepackage{amsfonts} 
\usepackage{epsfig} 
\usepackage{epsf} 
\usepackage{array}
\usepackage{color}

\begin{document}

\title{Microscopic Inhomogeneity and Superconducting Properties of a Two-dimensional Hubbard Model for High-$T_c$ Cuprates} 
\author{Satoshi Okamoto$^1$}

\author{Thomas A. Maier$^2$} \affiliation{$^1$Materials Science and Technology
Division, Oak Ridge National Laboratory, Oak Ridge, Tennessee 37831-6071, USA
\\ $^2$Computer Science and Mathematics Division and Center for Nanophase
Materials Sciences, Oak Ridge National Laboratory, Oak Ridge, Tennessee
37831-6494, USA}

\begin{abstract}
Recent scanning tunneling microscopy measurements on cuprate
superconductors have revealed remarkable spatial inhomogeneities in
the single-particle energy gap. Using cellular dynamical mean-field
theory, we study the zero temperature superconducting properties of
a single-band Hubbard model with a spatial modulation of the
electron density. We find that the inhomogeneity in the electronic
structure results in a substantial spatial variation in the
superconducting order parameter and single-particle energy gap,
reminiscent of the experimental results. In particular, we find that
the order parameter and gap amplitudes in the hole-rich regions are
significantly enhanced over the corresponding quantities in a
uniform system, if the hole-rich regions are embedded in regions
with smaller hole density.
\end{abstract}

\maketitle

\section{Introduction}
The field of strongly correlated electron systems has been one of the most
active areas of materials science for decades.\cite{Imada98} In particular,
transition-metal oxides including high-$T_c$ cuprates\cite{Bednorz96} and
colossal magnetoresistance manganites are fascinating both from fundamental
science and practical application points of view. Recent experimental studies revealed
that various transition-metal oxides exhibit spatial inhomogeneity in the
electronic structure.\cite{Dagotto05} 
For the cuprates, such inhomogeneity has been reported to
have random, checkerboard or stripe-type spatial
modulation.\cite{Lang02,Vershinin04,Hanaguri04,Kohsaka07} 

Although its origin and physical properties remain a center of debate, 
it is clear that spatial inhomogeneity in the
cuprates greatly affects the superconducting properties.\cite{McEloy05,Gomes07,Pasupathy08,Slezak08,Parker10} 
One of the most intriguing experimental observations is the positive correlation between the
superconducting gap and the oxygen dopant atoms \cite{McEloy05} and the
anticorrelation with the distance between a Cu atom and an apical O atom,
$d_A$, in the scanning tunneling microscopy (STM) experiments.\cite{Slezak08} 
Based on an early mean-field study of an inhomogeneous $tJ$ model, Wang {\it et
al}.\cite{Wang02} conjectured that the inhomogeneity arises from variations in
the local oxygen dopant level. Other papers have addressed the inhomogeneity by
introducing variations in coupling constants on a phenomenological mean-field
level.\cite{Andersen07,Yang07}  More recently, Mori {\it et al.}\cite{Mori08} proposed a
link between the variation in pairing interactions and the position of
out-of-plane oxygen atoms, that involves the covalency between the in-plane
$d_{x^2-y^2}$ band and the apical $p$ orbitals, and the screening of the
in-plane Coulomb repulsion on the Cu orbitals by the apical oxygen polarization. 
Here we study this problem in a microscopic model with inhomogeneity in the electronic structure 
using a non-perturbative approach that goes beyond a static mean-field treatment. 

In particular, we investigate the effect of spatial inhomogeneity in the
electronic structure on the superconducting (SC) properties of a Hubbard model 
using cellular dynamical mean-field theory (CDMFT).\cite{Kotliar01,Kotliar06}
Motivated by the result in Wang {\it et al.} \cite{Wang02} and Mori {\it et
al.} \cite{Mori08} where out-of plane dopant atoms affect the in-plane hole
density, we consider a two-dimensional Hubbard model with planar density
modulations. We find that the $d$-wave gap amplitude is significantly increased
in regions with increased hole density, if they are surrounded by regions with
smaller hole density. The spatial modulation of the gap is found to become as
large as a factor 2 and goes well beyond the variation one would expect from
the difference in hole density. The SC order parameter also shows similar
variation. The enhanced $d$-wave gap in the overdoped region is consistent with
the experimental reports if the main effect of oxygen atoms is to locally
increase the hole concentration.\cite{McEloy05,Slezak08} 

\section{Model and method}
\begin{figure} [tbp]
  \includegraphics[width=0.7\columnwidth,clip]{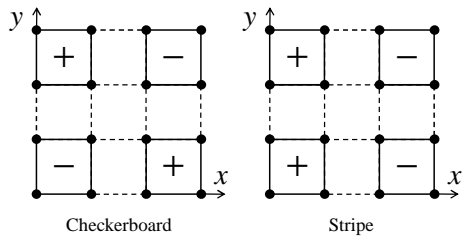}
  \caption{Schematics of the potential modulation. Left: checkerboard-type
  modulation and right: stripe-type modulation. On the sites belonging to a $2
  \times 2$ plaquette denoted by $+(-)$, the on-site potential is $+V$ ($-V$).
  } \label{fig:potential} 
\end{figure}

We consider a two-dimensional (2D) Hubbard model for the Cu $d$ electrons: 
\begin{equation} H = \sum_{i, \sigma} \varepsilon_i n_{i \sigma} - t
  \sum_{\langle ij \rangle, \sigma} \bigl(d_{i \sigma}^\dag d_{j \sigma} + H.c.
  \bigr) + U \sum_i n_{i \uparrow} n_{i \downarrow}. \label{eq:Hlatt} 
\end{equation}
Here, $d_{i \sigma}$ is the annihilation operator of an electron with spin
$\sigma$ at site $i$, $n_{i \sigma} = d_{i \sigma}^\dag d_{i \sigma}$, $t$ the
nearest-neighbor transfer integral, and $U$ the on-site repulsive Coulomb
interaction. 
Here, only the nearest-neighbor hopping $t$ is included. 
$\varepsilon_i$ is a site-dependent on-site potential, leading to 
a spatial inhomogeneity in the hole density. 
We consider two types of potential
modulations as illustrated in Fig.~\ref{fig:potential}. A checkerboard-type
variation in the local potential where $\varepsilon_i$ alternates between
$\varepsilon_i=+V$ and $-V$ on 2$\times$2 plaquettes and a stripe-like
variation in which the modulation is unidirectional.   

We study the zero temperature SC properties of the uniform and inhomogeneous
systems using the CDMFT. For a uniform system, the CDMFT maps the bulk lattice
onto a cluster embedded in a dynamic mean-field that describes the rest of the
system and is determined self-consistently. Here, we use a small 2$\times$2
cluster. This approximation may be justified in the strong-coupling $(U)$
limit, in which the coherence length becomes sufficiently small so that the
superconducting behavior is well approximated on a 2$\times$2 plaquette. In
addition, the small cluster size allows us to solve the effective cluster
problem using the Lanczos exact diagonalization (ED) technique
\cite{Caffarel94,Kancharla08} to directly access dynamical quantities such as
the density of states without additional analytic continuation. 

For the inhomogeneous system, we map the lattice onto a number  $N_c$ of in-equivalent
2$\times$2 clusters, indicated by the bold lines in Fig.~\ref{fig:potential},
where $N_c$ is the number of clusters per unit cell. 
In our case we have $N_c=2$. 
In this approximation, the self-energy depends on the cluster index
$R_I$ and the intracluster coordinates $\vec r_i$. 
This means that dynamic correlations are retained within a cluster, 
but neglected between the clusters.\cite{Potthoff99,Okamoto04b} 
The clusters are then linked by the self-consistency condition. 
This approach can be regarded as a cluster extension to the DMFT study of stripes in Ref.~\onlinecite{Fleck00}.
The choice of clusters may not be unique. 
For example, one may take clusters comprised of two sites of the $+V$ region and two sites of the $-V$ region. 
However, in this case, the local symmetry inside the cluster is severely broken. 
We expect this to give rise to a strong suppression of $d$-wave pairing inside the cluster. 
Hence this choice would not allow us to explore the physics we set out to explore, 
such as the modulation of a local pairing gap through the proximity to a region with different doping. 

With our choice of clusters, the impurity Hamiltonian is given by 
\begin{eqnarray} H_{imp,I} \!\!&=&\!\! H_I + \sum_{\alpha i \sigma}
  E^I_{\alpha} a_{\alpha i \sigma}^\dag a_{\alpha i \sigma} +\sum_{\alpha i
  \sigma} \Bigl\{ v^I_{\alpha} c_{i \sigma}^\dag a_{\alpha i \sigma} + h.c.
  \Bigr\} \nonumber \\ 
  && \!\! +  \sum_{\alpha \sigma} \Bigl\{ \Delta^I_{\alpha x}
  \sigma ( a_{\alpha 1 \sigma} a_{\alpha 2 - \sigma} + a_{\alpha 3 \sigma}
  a_{\alpha 4 - \sigma} ) \nonumber \\ 
  && \!\! +  \Delta^I_{\alpha y} \sigma
  (a_{\alpha 2 \sigma} a_{\alpha 3 -\sigma} + a_{\alpha 4 \sigma} a_{\alpha 1 -
  \sigma} ) + h.c. \Bigr\}. \label{eq:Himp} 
\end{eqnarray}
Here, $H_I$ represents the lattice Hamiltonian, Eq.~(\ref{eq:Hlatt}),
restricted to the $I$th region $(I=1, \ldots, N_c)$. We introduced two kinds of
bath orbitals $\alpha (=1,2)$, which describe the rest of the system and whose
energy level and hybridization strength with a cluster site are given by
$E_\alpha$ and $v_\alpha$, respectively. 
Thus, our cluster model consists of four interacting orbitals and eight bath orbitals. 
We also introduced two parameters
representing the superconducting correlations for bath $\alpha$,
$\Delta_{\alpha x}$ and $\Delta_{\alpha y}$. For the uniform system and the
checkerboard-type potential, the local tetragonal symmetry is retained and
therefore $\Delta_{\alpha y}=-\Delta_{\alpha x}$ for the $d$-wave pairing. On
the other hand in the stripe-type potential, the tetragonal symmetry is broken
and $\Delta_{\alpha x}$ and $\Delta_{\alpha y}$ are independent parameters.
Even in this case, the difference between $\Delta_{\alpha y}$ and
$-\Delta_{\alpha x}$ was found to be very small.

The self-consistency condition of the ED CDMFT is closed using the conjugate
gradient minimization algorithm with a distance function defined on an imaginary
frequency axis with the effective inverse temperature $\beta =50/t$, and short
cut-off frequency $\omega_c = 1.5 t$.\cite{Kancharla08} We set the Coulomb
repulsion to $U = 10t$ and note that our results are insensitive to this choice
as long as $U$ is sufficiently large. 
Potential modulation $V$ will be at most on the order of $t$, 
and Hubbard bands do not overlap with the low-energy quasiparticle part. 
Therefore, the effect of the short cut-off frequency is expected to be small. 

\begin{figure}[tbp] \includegraphics[width=0.7\columnwidth,clip]{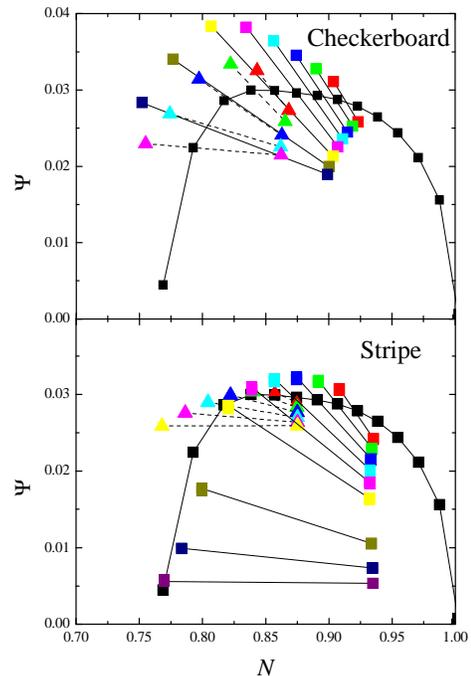}
  \caption{(Color online) The superconducting order parameter $\Psi$ as a
  function of carrier density $N$ for the 2D Hubbard model computed by CDMFT
  with the ED impurity solver at $T=0$. Black symbols are the results for the
  uniform systems. Colored (shaded) symbols connected by a thin line denote the
  results with a checkerboard-type potential modulation (upper panel) and a
  stripe-type potential modulation (lower panel). }
\label{fig:PsiED} 
\end{figure}

\section{Results}
Numerical results for the $d$-wave SC order parameter $\Psi = \Psi_x =-\Psi_y$
$(\Psi_\xi = \langle c_{i \uparrow} c_{i+\xi \downarrow} \rangle )$ as a
function of carrier concentration $N=\langle \sum_\sigma n_\sigma
\rangle_{cluster}$ for the uniform system are shown as black dots in
Fig.~\ref{fig:PsiED}. These results are in excellent agreement with the
previous work by Kancharla {\it et al.}\cite{Kancharla08} We find an optimal
carrier density $N^{opt} \sim 0.85$ which maximizes $\Psi$ and the
superconductivity vanishes for densities smaller than $N \sim 0.76$. 

We now turn to the results for the inhomogeneous systems for the checkerboard
and stripe-like modulations as shown in Fig.~\ref{fig:potential}. 
We plot the order parameter of cluster $I$, $\Psi^I$, at the corresponding densities
$N=N_I=\langle \sum_\sigma n_\sigma \rangle_{cluster,I}$ in
Fig.~\ref{fig:PsiED} as colored (shaded) symbols connected by thin lines. One
clearly observes that $\Psi$ in the cluster with lower electron filling (larger
hole density) is significantly enhanced over that in the uniform system at the
same filling. 
For a range of fillings, it becomes even larger than
the maximum value for the uniform system. At the same time, $\Psi$ in the
cluster with the larger filling is reduced relative to the homogeneous system.
Potential modulation, chemical potential, and the resulting carrier densities and SC order parameters are summarized in Table~\ref{tab:checkerboard}. 
Here, $I=+(-)$ corresponds to the cluster in which the potential is increased (decreased) by $V$, thus $N_+<N_-$. 
In the actual computations, we fix $\mu+V$ and vary $\mu-V$. 
One can also see that $V=0$ recovers the uniform system. 

\begin{table}[tbp]
\caption{Potential modulation $V$ for the checkerboard-type pattern, 
the resulting carrier densities $N_+$ and $N_-$, and superconducting order parameters $\Psi^\pm$. 
$\mu$ is the chemical potential. 
Results with $V=0$ correspond to the uniform system. 
In the actual computations, we fix $\mu+V$ and vary $\mu-V$. 
}
\label{tab:checkerboard}
\begin{center}
\begin{tabular}{cccccc} \hline \hline
$\mu/t$ & $V/t$ & \quad $N_+$ & \quad $N_-$ \quad & $\Psi^+$ & \quad $\Psi^-$ \\
\hline 
2.4  \quad & 0    & \quad 0.939 & \quad 0.939 & \quad 0.0265 & \quad 0.0265 \\
2.25 \quad & 0.15 & \quad 0.904 & \quad 0.923 & \quad 0.0311 & \quad 0.0258 \\
2.2  \quad & 0.2  & \quad 0.890 & \quad 0.919 & \quad 0.0328 & \quad 0.0252 \\
2.15 \quad & 0.25 & \quad 0.874 & \quad 0.915 & \quad 0.0345 & \quad 0.0245 \\
2.1  \quad & 0.3  & \quad 0.856 & \quad 0.911 & \quad 0.0364 & \quad 0.0236 \\
2.05 \quad & 0.35 & \quad 0.835 & \quad 0.907 & \quad 0.0382 & \quad 0.0226 \\
2.0  \quad & 0.4  & \quad 0.807 & \quad 0.903 & \quad 0.0384 & \quad 0.0213 \\
1.95 \quad & 0.45 & \quad 0.777 & \quad 0.900 & \quad 0.0340 & \quad 0.0200 \\
1.9  \quad & 0.5  & \quad 0.752 & \quad 0.899 & \quad 0.0284 & \quad 0.0190 \\
2.0  \quad & 0    & \quad 0.875 & \quad 0.875 & \quad 0.0296 & \quad 0.0296 \\
1.9  \quad & 0.1  & \quad 0.843 & \quad 0.868 & \quad 0.0325 & \quad 0.0273 \\
1.85 \quad & 0.15 & \quad 0.822 & \quad 0.865 & \quad 0.0334 & \quad 0.0259 \\
1.8  \quad & 0.2  & \quad 0.797 & \quad 0.863 & \quad 0.0314 & \quad 0.0241 \\
1.75 \quad & 0.25 & \quad 0.774 & \quad 0.862 & \quad 0.0269 & \quad 0.0226 \\
1.7  \quad & 0.3  & \quad 0.755 & \quad 0.862 & \quad 0.0230 & \quad 0.0215 \\
\hline \hline
\end{tabular}
\end{center}
\end{table}

\begin{table}[tbp]
\caption{Potential modulation $V$ for the stripe-type pattern, 
the resulting carrier densities $N_+$ and $N_-$, and superconducting order parameters $\Psi^\pm_{x(y)}$. 
$\mu$ is the chemical potential. 
}
\label{tab:stripe}
\begin{center}
\begin{tabular}{cccccccc} \hline \hline
$\mu/t$ & $V/t$ & \quad $N_+$ & \quad $N_-$ & \quad $\Psi^+_x$ & \quad $-\Psi^+_y$ & \quad $\Psi^-_x$ & \quad $-\Psi^-_y$ \\
\hline 
2.3  \quad & 0.1  & \quad 0.908 & \quad 0.935 & \quad 0.0307 & \quad 0.0305 & \quad 0.0241 & \quad 0.0243 \\
2.25 \quad & 0.15 & \quad 0.892 & \quad 0.934 & \quad 0.0318 & \quad 0.0316 & \quad 0.0228 & \quad 0.0230 \\
2.2  \quad & 0.2  & \quad 0.875 & \quad 0.933 & \quad 0.0323 & \quad 0.0319 & \quad 0.0214 & \quad 0.0216 \\
2.15 \quad & 0.25 & \quad 0.857 & \quad 0.933 & \quad 0.0320 & \quad 0.0317 & \quad 0.0200 & \quad 0.0201 \\
2.1  \quad & 0.3  & \quad 0.839 & \quad 0.932 & \quad 0.0310 & \quad 0.0306 & \quad 0.0183 & \quad 0.0185 \\
2.05 \quad & 0.35 & \quad 0.820 & \quad 0.932 & \quad 0.0286 & \quad 0.0281 & \quad 0.0163 & \quad 0.0165 \\
2.0  \quad & 0.4  & \quad 0.800 & \quad 0.934 & \quad 0.0177 & \quad 0.0174 & \quad 0.0105 & \quad 0.0106 \\
1.95 \quad & 0.45 & \quad 0.784 & \quad 0.935 & \quad 0.0099 & \quad 0.0099 & \quad 0.0073 & \quad 0.0074 \\
1.9  \quad & 0.5  & \quad 0.770 & \quad 0.935 & \quad 0.0056 & \quad 0.0058 & \quad 0.0053 & \quad 0.0053\\
1.95 \quad & 0.05 & \quad 0.857 & \quad 0.875 & \quad 0.0303 & \quad 0.0303 & \quad 0.0290 & \quad 0.0290 \\
1.9  \quad & 0.1  & \quad 0.840 & \quad 0.875 & \quad 0.0304 & \quad 0.0303 & \quad 0.0283 & \quad 0.0283 \\
1.85 \quad & 0.15 & \quad 0.822 & \quad 0.875 & \quad 0.0299 & \quad 0.0299 & \quad 0.0276 & \quad 0.0277 \\
1.8  \quad & 0.2  & \quad 0.804 & \quad 0.875 & \quad 0.0290 & \quad 0.0289 & \quad 0.0270 & \quad 0.0270 \\
1.75 \quad & 0.25 & \quad 0.786 & \quad 0.875 & \quad 0.0276 & \quad 0.0275 & \quad 0.0263 & \quad 0.0264 \\
1.7  \quad & 0.3  & \quad 0.768 & \quad 0.875 & \quad 0.0259 & \quad 0.0259 & \quad 0.0259 & \quad 0.0259 \\
\hline \hline
\end{tabular}
\end{center}
\end{table}

This enhancement of superconductivity is similar to that observed in our
previous study of superlattices made up of underdoped and overdoped Hubbard
layers.\cite{Okamoto08} There we argued that the enhancement in the overdoped
layers was due to their proximity to the larger pairing scale in the underdoped
layers. The similarity between the two results and the nature of approximation
lead us to speculate that the origin of the increased order parameter in the
present study has similar origin.

To support this conjecture, we plot the results for the system with the
stripe-like modulation in the lower panel of Fig.~\ref{fig:PsiED}. 
The corresponding numerical data are listed in Table~\ref{tab:stripe}. 
Here, we
plot both $\Psi_x$ and $-\Psi_y$. However, the difference between them is at
most $\sim 1\%$ and therefore the corresponding points are almost
indistinguishable (see also Table~\ref{tab:stripe}). 
As in the checkerboard case, the order parameter is enhanced
in the cluster with larger hole doping. However, it is clear that the
enhancement is significantly weaker than in the former case with
checkerboard-type potential. For the checkerboard modulation, the maximum
$\Psi$ is about 30 \% larger than the maximum value in the uniform system,
while in the striped case the enhancement is about 10 \%. We believe that this
difference comes from the fact that, for the checkerboard modulation, 
the hole-rich cluster is fully surrounded by hole-poor clusters, which have a larger
pairing scale.\cite{Maier06}  On the other hand, in the striped case, the
hole-rich cluster is in proximity to only two hole-poor clusters and also two
hole-rich clusters, which reduces the effect. In both cases, $\Psi$ in the
hole-poor clusters is decreased significantly by the coupling with the
hole-rich clusters. The reduction is stronger in the stripe-type potential. 
This fact may partly be explained by the breaking of the local tetragonal symmetry. 
These results suggest that the SC order parameter is quite sensitive to the
local environment.

It is also interesting to discuss the dependence of the order parameters in the hole-rich, $\Psi^+$, 
and hole-poor regions, $\Psi^-$, with increasing potential $V$, 
or equivalently, decreasing filling $N_+$ and $N_-$, respectively. 
With increasing $V$ and decreasing $N_+$, $\Psi^+$ follows the same trend as the order parameter 
in the homogeneous system, i.e., it first increases, then decreases with decreasing $N_+$. 
It is enhanced over the order parameter in the uniform system through the proximity of the hole-rich region 
to the larger pairing scale in the hole-poor region.  
An optimal value of $V$ that maximizes $\Psi^+$ results from a delicate balance 
between the proximity of the larger pairing scale from the hole-poor regions, 
and the reduction in the pairing scale with increasing doping in its own hole-rich region. 
The order parameter $\Psi^-$ in the hole-poor region, in contrast, 
decreases monotonically with decreasing filling $N^-$. 
This may be explained by a strong reduction in the pairing interaction in the hole-poor region 
by its coupling to the smaller pairing scale in the hole-rich regions.

It would be interesting to determine whether the enhancement of pairing we find in the hole-rich regions, 
translates to an increase in $T_c$. 
Our zero temperature and small cluster calculations, however, do not allow us to speculate on this question. 
Finite temperature calculations on larger clusters similar to those carried out 
for a striped Hubbard model in Ref.~\onlinecite{Maier10} will be necessary to gain more insight into this question. 
Work along these lines is currently underway.

\begin{figure} [tbp] \includegraphics[width=0.8\columnwidth,clip]{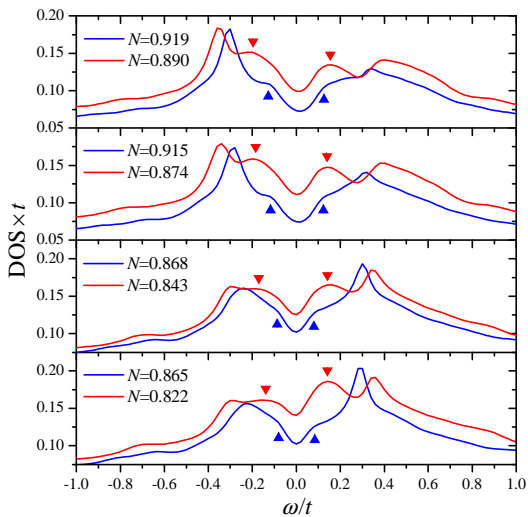}
  \caption{(Color online) Local DOS of the Hubbard model with the
  checkerboard-type potential modulation. Each panel shows the local DOS in the
  clusters with smaller $N$ and larger $N$ for a given potential modulation.
  Down (up) triangles indicate the low-energy structure for the smaller
  (larger) $N$ region. To compute the self-energy, a small imaginary number
  $\eta=0.125 t$ was added to the real frequency $\omega$.} \label{fig:dos} 
\end{figure}

From the variation in the SC order parameter, one may expect that the $d$-wave
gap in the density of states (DOS) is also modulated in the inhomogeneous
systems. To see this effect, we computed the local DOS for different strengths
of the checkerboard-type potential modulations. We calculated the $\vec
k$-space average of the momentum-dependent lattice Green's function to obtain
the local Green's function, and the local DOS is given as its imaginary part.
The lattice Green's function has the same form as that used to close the
self-consistency condition for the CDMFT but with the self-energy as a function
of real frequency. 
When computing the self-energy, we introduced a small imaginary number $\eta$ on the order of $0.1t$ 
to the real frequency. 
We confirmed that different choices of $\eta$ give similar results. 
Results presented below were obtained using $\eta=0.125 t$. 
Because of the finite number of orbitals in the ED CDMFT and finite broadening $\eta$, we cannot
access the very low-energy behavior. Yet, we believe that the main structure of
the local DOS is accessible. 

Results of the local DOS are shown in Fig.~\ref{fig:dos}. Each panel
corresponds to a different strength of potential modulation, and the resulting
electron densities in hole-rich and hole-poor clusters are indicated. 
One can see that the low-energy structure (peak or shoulder) for the smaller $N$
region is located at higher energy than for the larger $N$ region. 
%

\begin{figure} [tbp] \includegraphics[width=0.7\columnwidth,clip]{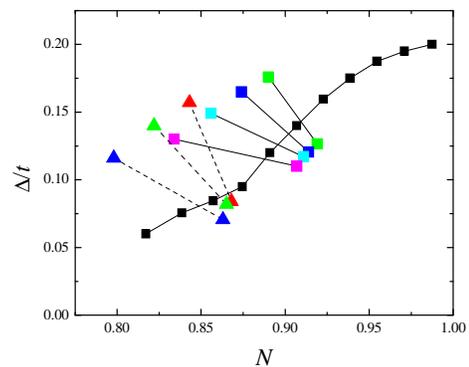}
  \caption{(Color online) $d$-wave gap amplitude $\Delta$ as a function of the
  average carrier density $N$ for the 2D Hubbard model computed by the CDMFT
  with the ED impurity solver at $T=0$. Black symbols are the results for the
  uniform systems. Colored (shaded) symbols connected by a thin line denote the
  results with a checkerboard-type potential modulation. } \label{fig:gap} 
\end{figure}

In order to see the dependence of the gap amplitude $\Delta$ on the carrier
density modulation more clearly, we plot $\Delta$ as a function of the average
density $N$ in Fig.~\ref{fig:gap} for the checkerboard-type potential
modulations. Here, we defined $\Delta$ as the half distance between the lowest
energy peaks or shoulders above and below $\omega=0$. For comparison, we also
plot $\Delta$ for the uniform system as black dots. The monotonic increase in
$\Delta$ with approaching $N=1$ is consistent with the previous
report.\cite{Kancharla08} From this figure, one clearly sees that the gap
amplitude in the hole-rich region is significantly enhanced over that in
hole-poor region, as well as that of the corresponding uniform system at the
same hole density.  
We stress here that this result is robust against the choice of broadening parameter $\eta$.

From Figs.~\ref{fig:dos} and \ref{fig:gap}, we also notice that the effect of a
potential modulation on the gap amplitude in the underdoped region is weaker
than that on $\Psi$; $\Delta$ is reduced less than $\Psi$ when the potential
modulation exists. We argue that this is caused by a near cancellation between
the reduced order parameter $\Psi$ and the enhanced quasiparticle weight $Z$
both due to the presence of the hole-rich region nearby as follows. 
The $d$-wave order parameter $\Psi$ is proportional to the anomalous self-energy at $\omega \rightarrow 0$, 
and the real part of the self-energy is given by $(1-1/Z)\omega$. 
The combination between the two effects leads to the gap amplitude $\Delta \propto Z \Psi$. 
This insensitivity may become important to understand the doping-dependent gap amplitude in the
presence of inhomogeneity. 

We emphasize that the enhancement of the $d$-wave gap in the hole-rich region
relative to that in the hole-poor region is quite striking and opposite to the
behavior in the uniform system. 
It is also contrary to the previous theoretical result based on a static mean-field theory.\cite{Wang02} 

\section{Discussion and summary}

Finally, let us briefly discuss the possible implication of the present study 
on the STM results. 
In Ref.~\onlinecite{McEloy05}, it was reported that the gap amplitude and the
position of dopant oxygen atoms have positive correlation. Furthermore, in
Ref.~\onlinecite{Slezak08}, an anticorrelation was reported between the gap
magnitude and the distance between the Cu and apical O, $d_A$. If the main
effect of the dopant oxygen and shorter $d_A$ is to increase the local hole
concentration, as reported in the calculation by Mori {\it et
al.},\cite{Mori08} our observation that the region with larger hole density
(smaller filling $N$) has a larger gap is consistent with the experimental
reports.

Estimating potential modulation from experimental data is not easy because 
it requires precise position of charged (dopant) ions and the electron-density modulation. 
But, taking the experimental uncertainty of Ref.\onlinecite{McEloy05}, 
density modulation is expected to be less than 10~\%. 
In our calculations, the potential modulation $V < t$ is enough to induce such a small density modulation. 
This $V$ ($< 0.5$~eV) is smaller than the one simply estimated by assuming point dopant-charged ions. 
For example, in La$_{2-x}$Sr$_x$CuO$_4$, potential modulation due to the A-site dopant ions becomes 
of order of 1--2~eV with the in-plane lattice constant 4~\AA \, and the dielectric constant $\varepsilon \sim 3.5$.%
\cite{Ohta91} 
In reality, other effects such as the screening due to the other conduction electrons reduce the modulation. 
Therefore, our potential modulation must be regarded as the self-consistently determined one 
by including screening effects, and $V \alt t$ seems to be in the realistic range. 
We would like to emphasize that our results show that even a very weak density modulation, 
on the order of 5~\% and therefore well within the experimental uncertainty, 
is enough to result in a significant, i.e., factor 2, variation in the local gap amplitude.

However, the variation in the gap magnitude we observe of about a factor of two is less
than the factor 3 that has been reported in Refs.~\onlinecite{McEloy05}, and \onlinecite{Slezak08}. 
Certainly, a single-band Hubbard model is too simple to fully resolve this issue, 
and additional degrees of freedom could play an important role, 
such as the apical oxygen as proposed in Ref.~\onlinecite{Mori08}. 
One possible scenario to resolve the present discrepancy is 
the screening effect by the oxygen ions. 
This reduces the on-site Coulomb interaction, resulting in a further enhancement of 
the superconducting order parameter $\Psi$ and the $d$-wave gap within the strong-coupling regime.\cite{Kancharla08}

As reported in the STM measurements in Ref.~\onlinecite{Gomes07}, both the
average gap amplitude and the width and asymmetry of the distribution of gap
magnitudes increase as the doping decreases, with long tails on the large gap
side of the distribution. Assuming that the observed inhomogeneity is related
to a carrier-density modulation, our study can provide a possible explanation. 
Based on our results, we can speculate that the large gap tail in the
underdoped samples originates from hole-rich islands with large gaps residing
in the hole-poor environment. The strong asymmetry is naturally expected
because the volume fraction of the large-gap hole-rich regions is small because
of the small average hole concentration.

Recently, Parker and coworkers\cite{Parker10} studied the gap-amplitude modulation 
in the over-doped cuprates Bi$_2$Sr$_2$CaCu$_2$O$_{8+\delta}$ as a function of temperature. 
It was revealed that medium-sized gap regions surrounded by large-gap regions 
persist to higher temperatures than 
regions with similar gap sizes in the proximity of small-gap regions. 
We speculate that both the large-gap and medium-gap regions are over-doped 
and the former is closer to the optimal doping. 
Thus, the proximity effect is weaker compared with that between 
under-doped regions and over-doped regions as we studied in this paper. 
To confirm this conjecture, 
it is necessary to make a link between the gap amplitude and the local carrier density. 
It is also desirable to perform the similar experimental analysis 
using the under-doped and near-optimally-doped cuprates.

To summarize, we have studied the superconducting properties of an
inhomogeneous two-dimensional Hubbard model for the high-$T_c$ cuprates with
modulations in the electronic hole density using a cellular dynamical mean-field approximation. 
We have found that the modulation in hole density results
in a considerable spatial variation in the superconducting order parameter
which is quite sensitive to the local environment. In particular, the order
parameter in the hole-rich regions is significantly enhanced over the order
parameter in a corresponding uniform system, if the hole-rich region is
embedded in a region with smaller hole-density.  Furthermore, we have found
that the $d$-wave gap amplitude in the density of states shows similar
modulation, with an enhanced gap in the hole-rich regions.  These results are
reminiscent of the variation in gap amplitudes reported in recent STM
measurements of cuprate superconductors.

\acknowledgements
The authors would like to thank D. J. Scalapino, J. C. Davis, and A. Yazdani for valuable discussions.  
This work was supported by the Materials Sciences and Engineering Division,
Office of Basic Energy Sciences, U.S. Department of Energy. A portion of this
research at Oak Ridge National Laboratory's Center for Nanophase Materials
Sciences was sponsored by the Scientific User Facilities Division, Office of
Basic Energy Sciences, U.S. Department of Energy.


\begin{thebibliography}{99}

\bibitem{Imada98}M. Imada, A. Fujimori, and Y. Tokura, Rev. Mod. Phys. {\bf 70}, 1039 (1998). 

\bibitem{Bednorz96}J. G. Bednorz and K. A. M{\"u}ller, Z. Phys. B {\bf 64}, 189 (1986).

\bibitem{Dagotto05}E. Dagotto, Science {\bf 309}, 257 (2005). 

\bibitem{Lang02} K. M. Lang, V. Madhavan, J. E. Hoffman, E. W. Hudson, H. Eisaki, S. Uchida, and J. C. Davis, Nature (London) {\bf 415}, 412 (2002).

\bibitem{Vershinin04}M. Vershinin, S. Misra, S. Ono, Y. Abe, Y. Ando, and A. Yazdani, Science {\bf 303}, 1995 (2004).

\bibitem{Hanaguri04}T. Hanaguri, C. Lupien, Y. Kohsaka, D.-H. Lee, M. Azuma, M. Takano, H. Takagi, J. C. Davis, Nature (London) {\bf 430}, 1001 (2004). 

\bibitem{Kohsaka07}Y. Kohsaka, C. Taylor, K. Fujita, A. Schmidt, C. Lupien, T. Hanaguri, M. Azuma, M. Takano, H. Eisaki, H. Takagi, S. Uchida, J. C. Davis, Science {\bf 315}, 1380 (2007). 

\bibitem{McEloy05}K. McElroy, J. Lee, J. A. Slezak, D.-H. Lee, H. Eisaki, S. Uchida, J. C. Davis, Science {\bf 309}, 1048 (2005). 

\bibitem{Gomes07}K. K. Gomes, A. N. Pasupathy, A. Pushp, S. Ono, Y. Ando, A. Yazdani, Nature (London) {\bf 447}, 569 (2007). 
\bibitem{Pasupathy08}A. N. Pasupathy, A. Pushp, K. K. Gomes, C. V. Parker, J. Wen, Z. Xu, G. Gu, S. Ono, Y. Ando, A. Yazdani, Science {\bf 320}, 196 (2008). 

\bibitem{Slezak08}J. A. Slezak, J. Lee, M. Wang, K. McElroy, K. Fujita, B. M. Andersen, P. J. Hirschfeld, H. Eisaki, S. Uchida, and J. C. Davis, Proc. Natl. Acad. Sci. U.S.A. {\bf 105}, 3203 (2008).

\bibitem{Parker10}C. V. Parker, A. Pushp, A. N. Pasupathy, K. K. Gomes, J. Wen, Z. Xu, 
S. Ono, G. Gu, and A. Yazdan, Phys. Rev. Lett. {\bf 104}, 117001 (2010). 

\bibitem{Wang02}Z. Wang, J. R. Engelbrecht, S. Wang, H. Ding, and S. H. Pan, Phys. Rev. B {\bf 65}, 064509 (2002). 

\bibitem{Andersen07}B. M. Andersen, P. J. Hirschfeld, J. A. Slezak, Phys. Rev. B {\bf 76}, 020507(R) (2007). 

\bibitem{Yang07}K.-Y. Yang, T. M. Rice, and F-C. Zhang, Phys. Rev. B {\bf 76}, 100501(R) (2007). 

\bibitem{Mori08}M. Mori, G. Khaliullin, T. Tohyama, and S. Maekawa, Phys. Rev. Lett. {\bf 101}, 247003 (2008). 


\bibitem{Kotliar01}G. Kotliar, S. Y. Savrasov, G. P{\'a}lsson, and G. Biroli, Phys. Rev. Lett. {\bf 87}, 186401 (2001).

\bibitem{Kotliar06} G. Kotliar, S. Y. Savrasov, K. Haule, V. S. Oudovenko, O. Parcollet, C. A. Marianetti, Rev. Mod. Phys. {\bf 78}, 865 (2006). 

\bibitem{Caffarel94}M. Caffarel and W. Krauth, Phys. Rev. Lett. {\bf 72}, 1545 (1994). 
\bibitem{Kancharla08}S. S. Kancharla, B. Kyung, D. S{\'e}n{\'e}chal, M. Civelli, M. Capone, G. Kotliar, A.-M. S. Tremblay, Phys. Rev. B {\bf 77}, 184516 (2008). 


\bibitem{Potthoff99}M. Potthoff and W. Nolting, Phys. Rev. B {\bf 59}, 2549 (1999).  
\bibitem{Okamoto04b}S. Okamoto and A. J. Millis, Phys. Rev. B {\bf 70}, 241104(R) (2004).  
\bibitem{Fleck00}M. Fleck, A. I. Lichtenstein, E. Pavarini, and A. M. Ole{\'s}, Phys. Rev. Lett. {\bf 84}, 4962 (2000).

\bibitem{Okamoto08}S. Okamoto and T. A. Maier, Phys. Rev. Lett. {\bf 101}, 156401 (2008). 

\bibitem{Maier06}T. A. Maier, M. Jarrell, and D. J. Scalapino, Phys. Rev. B {\bf 74}, 094513 (2006). 

\bibitem{Maier10}T. A. Maier, G. Alvarez, M. Summers, and T. C. Schulthess, Phys. Rev. Lett. {\bf 104}, 247001 (2010). 

\bibitem{Ohta91}Y. Ohta, T. Tohyama, and S. Maekawa, Phys. Rev. B {\bf 43}, 2968 (1991). 


\end{thebibliography}
\end{document}